\documentclass[
prc,%
10pt,%
final,%
notitlepage,%
oneside,%
twocolumn,%
nobibnotes,%
nofootinbib,% In the current version of REVTeX, when this option is on,
superscriptaddress,%
floatfix,%
showkeys,%
showpacs]%
{revtex4}
\usepackage{color}
\usepackage{amsfonts}
\usepackage{amsbsy}
\usepackage{mathrsfs}
\usepackage{graphicx}           % include figures
\def\lsim{\mathrel{\rlap{
\lower4pt\hbox{\hskip-3pt$\sim$}}
    \raise1pt\hbox{$<$}}}     %less than approx. symbol
\def\gsim{\mathrel{\rlap{
\lower4pt\hbox{\hskip-3pt$\sim$}}
    \raise1pt\hbox{$>$}}}     %greater than or approx. symbol
\def\scr#1{\mbox{\scriptsize #1}}

\begin{document}
%\today

\title{Directed Flow Indicates a Crossover Deconfinement Transition
in Relativistic Nuclear Collisions}
%of Protons, Antiprotons and Pions in Heavy-Ion Collisions at $\sqrt{s_{NN}}$ from 2.7 to 27 GeV}

\author{Yu. B. Ivanov}\thanks{e-mail: Y.Ivanov@gsi.de}
\affiliation{National Research Centre "Kurchatov Institute" (NRC "Kurchatov Institute"), 123182 Moscow, Russia}
\affiliation{National Research Nuclear University "MEPhI" (Moscow Engineering
Physics Institute), 115409 Moscow, Russia}
\author{A. A. Soldatov}\thanks{e-mail: saa@ru.net}
\affiliation{National Research Nuclear University "MEPhI" (Moscow Engineering
Physics Institute),
%Kashirskoe sh. 31, 
115409 Moscow, Russia}
%\affiliation{Moscow State University, 
%Moscow, Russia}

\begin{abstract}
Analysis of directed flow ($v_1$) of protons,  antiprotons and pions in heavy-ion collisions
is performed   in the range of incident energies  $\sqrt{s_{NN}}$ = 2.7--27 GeV. 
Simulations have been done within a three-fluid model
employing a purely hadronic equation of state (EoS) and two versions of the EoS
involving deconfinement transitions: 
a first-order phase transition and a smooth crossover transition.
High sensitivity of the directed flow, especially the proton one, to the EoS is found.
The crossover EoS is favored
by the most part of considered experimental data. 
A strong wiggle in the excitation function of the proton $v_1$ slope  
at the midrapidity obtained with the first-order-phase-transition EoS
and a smooth proton $v_1$ with positive midrapidity slope, within the hadronic EoS 
unambiguously disagree with the data. 
The pion and antiproton $v_1$ also definitely testify in favor of the crossover EoS. 
The results obtained with deconfinement EoS's apparently 
indicate that these  EoS's in the quark-gluon sector should be 
stiffer at high baryon densities than those used in the calculation. 
\end{abstract}

%\pacs{25.75.-q, 24.85.+p, 12.38.Mh}

\maketitle

%______________________________________________________________________
\section{Introduction}

The directed flow \cite{Danielewicz:1985hn} of particles has been one of the key
observables, since first data on the heavy ion collisions
became available at the Bevalac. Nowadays it is defined as  
the first  coefficient, $v_1$, in the Fourier expansion of a particle distribution, ${d^2 N}/{d y\;d\phi}$,  
in azimuthal angle $\phi$ with respect to the reaction plane \cite{Voloshin:1994mz,Voloshin:2008dg}
\begin{eqnarray}
 \label{vn-def}
\frac{d^2 N}{d y\;d\phi} = \frac{d N}{dy}
\left(1+ \sum_{n=1}^{\infty} 2\; v_n(y) \cos(n\phi)\right),
\end{eqnarray}
where $y$ is a longitudinal rapidity of a particle. 
The directed flow is mainly formed at an early
(compression) stage of the collisions and hence is sensitive to
early pressure gradients in the evolving nuclear matter \cite{So97,HWW99}. The
harder EoS is, the stronger pressure is developed.
Thus the flow reflects the stiffness of the nuclear EoS 
at the early stage of nuclear collisions \cite{Danielewicz:2002pu,RI06},
which is of prime interest for heavy-ion research. 
A retrospective review and a survey of new developments in the field of the collective flow 
is presented in recent article  \cite{Csernai:2014cwa}.

The directed flow has been extensively exploited to
obtain information on the EoS. 
In particular, it
was predicted that the first-order transition to the quark-gluon phase (QGP)  
results in significant reduction of the directed flow \cite{HS94,Ri95,Ri96}
(the so-called "softest-point" effect), because the pressure gradients in the
mixed phase are lower than those in pure hadronic and quark-gluon phases.
The $v_1$ data \cite{E877,E895} from the BNL Alternating Gradient Synchrontron (AGS)   
indeed demonstrate a graduate fall of the slope
of $v_1(y)$ at midrapidity with the incident energy rise from $\sqrt{s_{NN}}$ = 2.7 GeV ($E_{\rm lab}=$ 2 $A\cdot$GeV)
to 4.3 GeV ($E_{\rm lab}=$ 8 $A\cdot$GeV). 
This finding was further developed within fluid-dynamical models \cite{CR99,Br00}. 
It was found that the directed flow as a function of rapidity exhibits a wiggle near the midrapidity 
with a negative slope near the midrapidity, when the incident energy is in the range  
corresponding to onset of the first-order phase transition. 
This occurs because the event shape at these energies resembles an ellipsoid
in coordinate space, tilted with respect to
the beam axis. This ellipsoid expands predominantly orthogonal
to its short dimension, forming
a so-called "third component" \cite{CR99} or "antiflow" \cite{Br00}  near the midrapidity. 
When the softest point is passed, i.e. the incident energy is above that 
corresponding to the onset the first-order phase transition, the  midrapidity $v_1$ 
slope reaches a maximum. After that the $v_1$ slope decreases again \cite{Ri95,Ri96,Br00,St05}. 
Thus, the  wiggle near the midrapidity and the wiggle-like behavior 
of the excitation function of the midrapidity $v_1$ slope were put forward as 
a signature of the QGP phase transition. 
Measurements of the directed flow by the NA49 collaboration \cite{NA49} at the CERN Super Proton Synchrotron (SPS)  
had insufficient statistics
to draw definite conclusions on presence or absence of such a $v_1$ wiggle at the midrapidity.

However, the midrapidity $v_1$ wiggle can have a different physical origin. 
The QGP EoS is not a necessarily
prerequisite to reach the stopping needed to create this
tilted source \cite{SSV00}. 
A combination of space-momentum correlations---characteristic
of radial expansion together with the correlation between the
position of a nucleon in the fireball and its stopping---may result
in a negative slope in the rapidity dependence of the directed flow
in high-energy nucleus-nucleus collisions.

The elliptic flow,  $v_2$, 
 and the triangular flow, $v_3$, have been extensively
studied both theoretically and experimentally in the last years by
about five orders of magnitude in the collision energy
$\sqrt{s_{NN}}$~\cite{STAR10}. In contrast, apart from first
measurements and till recent times, the directed
flow was insufficiently experimentally studied  to check the above predictions.
The interest in the directed flow has recently been revived due to new 
data obtained by the STAR collaboration within the framework of the beam energy scan (BES) program 
at the BNL Relativistic Heavy Ion Collider (RHIC)\cite{STAR-14}. 
The directed flow of
identified hadrons---protons, antiprotons, positive and negative pions---has been measured with high
precision for  Au+Au collisions in the energy range $\sqrt{s_{NN}}$ =(7.7-39) GeV. These data
together with earlier experimental results from the AGS \cite{E877,E895}  and SPS \cite{NA49}
provide a basis for theoretical analysis of the directed flow in a wide energy range.

These data  have been already addressed in Refs. \cite{SAP14,Konchakovski:2014gda}. 
The Frankfurt group \cite{SAP14} confined itself to incident energies $\sqrt{s_{NN}}<$ 20 GeV. 
However, the authors of Ref. \cite{SAP14} did not succeed to describe
the data and to obtain conclusive results. Within a
hybrid approach \cite{Petersen:2008dd}, the authors found that there is no sensitivity of 
the directed flow on the EoS and,
in particular, on the existence of a first-order phase transition.
The reason of this result can be that 
the initial interpenetration stage of the collision 
is described within the Ultrarelativistic Quantum Molecular
Dynamics (UrQMD)\cite{Bass98}  for all scenarios (with and without transition to the QGP) in the
hybrid approach \cite{Petersen:2008dd}. 
Because of the UrQMD model the effective EoS during this stage is purely hadronic.
Only later, when transition from initial UrQMD
transport to the fluid dynamics happens, different scenarios
start to differ. 
As mentioned above, the directed flow is mainly formed at the early
stage of the collisions \cite{So97,HWW99}.
Therefore, in all scenarios considered in the
hybrid approach \cite{Petersen:2008dd} the directed flow was mainly formed
at the purely hadronic UrQMD stage, thus exhibiting similar results 
for different scenarios.

In Ref. \cite{Konchakovski:2014gda}
the new STAR data were analyzed within 
two complementary approaches: kinetic transport approaches
of the parton-hadron string dynamics (PHSD) \cite{CB09}
and the hadron string dynamics (HSD) \cite{PhysRep}),
and a hydrodynamic approach of the relativistic three-fluid dynamics (3FD) \cite{IRT06,Iv13-alt1}.
The PHSD model includes a crossover-type transition into the QGP, while 
the HSD one is a purely hadronic version of the PHSD. 
The 3FD simulations were preformed with two EoS's: a purely hadronic
EoS \cite{GM79} and a EoS with a crossover transition into the QGP \cite{KRST06}. 
It was found that the directed flow is sensitive to the EoS. 
The crossover scenario within both the PHSD and 3FD provides the best (but not perfect) results
being in a reasonable agreement with the STAR data. 

In the present paper we extend the analysis performed in Ref. \cite{Konchakovski:2014gda} 
within the 3FD model. Results for a EoS with a first-order phase transition \cite{KRST06} are 
reported. The AGS \cite{E895} and SPS \cite{NA49} data are considered in detail
on equal footing with the new STAR results. 
Computations are performed with somewhat higher accuracy, i.e. a finer grid and larger numbers 
of test particles% 
\footnote{A numerical "particle-in-cell" scheme is used in the
  present simulations, see Ref.~\cite{IRT06} and references therein
  for more details. The matter transfer due to pressure gradients and 
  friction between fluids is
  computed on a fixed grid (so called Euler step of the scheme). An
  ensemble of Lagrangian test particles is used for the calculation of
  the drift transfer of the baryonic charge, energy, and momentum (so
  called Lagrangian step of the scheme).}, 
even than that in Ref. \cite{Konchakovski:2014gda}. 
In contrast to other observables, the
directed flow is very sensitive to the accuracy settings of the numerical scheme. 
Accurate calculations require a very high memory and computation time. 
In particular, due to this reason we failed 
to perform calculations for energies above $\sqrt{s_{NN}}=$ 30
GeV. Note that the change of other observables, analyzed so
far~\cite{Iv13-alt1,Iv13-alt2,Iv13-alt3,Ivanov:2012bh,Ivanov:2013cba,Ivanov:2013mxa,Iv14}, 
due to higher accuracy is below 15\% as
compared to results of previous calculations.

\section{The 3FD model}
\label{sec:3FD}

The 3FD model~\cite{IRT06} is an  extension of a
two-fluid model with radiation of direct
pions~\cite{MRS88,RIPH94,MRS91} and a (2+1)-fluid
model~\cite{Kat93,Brac97}. These models have been further elaborated to include
a baryon-free (so called fireball) fluid on an equal footing with the baryon-rich ones. A
certain formation time was introduced for the fireball fluid,
during which the matter of the fluid propagates without
interactions. The formation time is associated with a finite
time of string formation and decay and is incorporated also in the
kinetic transport models such as PHSD/HSD \cite{CB09,PhysRep} and UrQMD \cite{Bass98}.
% (with $\tau \approx$ 0.8 fm/c).

The 3FD model~\cite{IRT06} describes a nuclear collision from the stage of 
the incident cold nuclei approaching each other, to the
final freeze-out stage. Contrary to the conventional one-fluid dynamics,
where a local instantaneous stopping of matter of the colliding nuclei
is assumed, the 3FD considers an inter-penetrating counter-streaming flows of leading baryon-rich
matter, which gradually decelerate each other due to mutual friction.
The basic idea of a 3FD approximation to heavy-ion
collisions~\cite{Iv87-1,Iv87-2} is that  a
generally nonequilibrium distribution of baryon-rich matter 
at each space-time point can be
represented as a sum of two distinct contributions initially
associated with constituent nucleons of the projectile and target
nuclei. In addition, newly produced particles, populating
predominantly the midrapidity region, are associated with the fireball
fluid. Therefore, the 3FD approximation is a minimal way to simulate
the early-stage nonequilibrium state of the colliding nuclei 
at high incident energies.

Friction forces between fluids are the key ingredients of the model that 
determine dynamics of the nuclear collision. 
The friction forces in the hadronic phase were estimated in Ref. \cite{Sat90}
based on experimental inclusive proton-proton cross sections. 
In order to reproduce the baryon stopping at
high incident energies, this estimated friction between counter-streaming
fluids was enhanced within the hadronic scenario \cite{Iv13-alt1}. 
%as compared to its estimate. 
Though such enhancement
is admissible in view of uncertainties of the estimated friction,
the value of the enhancement looks too high. In deconfinement scenarios 
there is no need to modify the hadronic friction \cite{Iv13-alt1}.
This can be considered as an indirect argument in favor of such
scenarios. 
At the same time, the friction forces in the QGP are purely phenomenological. 
They were fitted to reproduce the baryon stopping at
high incident energies within the deconfinement scenarios. 
There are no theoretical estimates of the QGP friction in terms of the QGP dynamics so far. 
A parton cascade model \cite{Xu:2004mz} offers a possible ground for such estimate. 
This approach predicts a fast kinetic equilibration (on a
scale of 1 fm/c) mainly driven  by the inelastic processes. 
Therefore, a reasonably strong QGP friction can be anticipated in this approach. 
Another promising way for such estimate can be based on an effective string rope model
\cite{Magas:2000jx,Magas:2002ge}. 
Proceeding from
coherent Yang-Mills field theoretical approach, this model introduces an effective
string tension based on Monte-Carlo string cascade and parton cascade
model results. The
effective string tension causes substantial baryon stopping in heavy-ion collisions.
In particular, for semi-central collisions this model predicts formation of a compact initial QGP fireball in the 
form of a tilted disk. Thus, it naturally explains the origin of the above-discussed 
``antiflow'' or ``third flow component''.

Different EoS's can be implemented in the 3FD model. 
All three fluids are described by the same EoS (chosen for the 
simulation), of course, with   
their specific values of the thermodynamic quantities. 
At the initial stage of the reaction 
all three fluids coexist in the same 
space-time region, thus describing a certain {\em nonequilibrium} state 
of the matter. 
It may happen that one or two of the fluids occur in the 
quark-gluon phase while other(s) is(are) in the hadronic one. This is a kind 
of a nonequilibrium mixed phase that is also possible in the model. 
A key point is that the 3FD model is able to treat a deconfinement transition at the
early {\em nonequilibrium} stage of the collision, when  
the directed flow is mainly formed, as it was mentioned above. 
This makes 3FD predictions for $v_1$, at least, sensitive to the used EoS.

In this work we apply a purely hadronic EoS~\cite{GM79}, an EoS with a
crossover transition as constructed in Ref.~\cite{KRST06}
and an EoS with a first-order phase transition into the QGP \cite{KRST06}. 
In recent  
works~\cite{Iv13-alt1,Iv13-alt2,Iv13-alt3,Ivanov:2012bh,Ivanov:2013cba,Ivanov:2013mxa,Iv14} an
analysis of the major part of bulk observables has been performed 
with these three EoS's:
the baryon stopping~\cite{Iv13-alt1,Ivanov:2013cba}, yields of different hadrons,
their rapidity and transverse momentum
distributions~\cite{Iv13-alt2,Iv13-alt3}, 
as well as the elliptic flow
excitation function~\cite{Ivanov:2013mxa,Iv14}.
%This analysis has been done in the
%same range of incident energies and within the same EoS's as those in the present paper. 
Comparison with available data, including those
at RHIC energies, indicated a definite advantage of the  
deconfinement (crossover and first-order) scenarios over the purely hadronic one 
especially at high collision energies.
However, predictions of the crossover and first-order-transition scenarios 
looked very similar so far.  
Only a slight preference could be given to the crossover EoS,  
though the latter 
%in the present version 
does not perfectly reproduced the data either.
The physical input
of the present 3FD calculations is described in detail in
Ref.~\cite{Iv13-alt1}. No tuning (or change) of physical 3FD-model parameters
has been done in the present study as compared to that stated in
Ref.~\cite{Iv13-alt1}.
A more detailed discussion of the features of the 3FD model can be 
found in Refs. \cite{IRT06,Iv13-alt1,Iv14}.

%______________________________________________________________________
\section{Directed flow within alternative scenarios}

As mentioned above, calculations of the directed flow require a
high numerical accuracy, i.e. a fine computational grid and a large number of test particles. 
This high accuracy is needed to accurately describe the initial stage of the collision, 
where  pressure gradients in the evolving nuclear matter are high. The accuracy at this stage is decisive for
the directed flow because the latter is mainly formed at the early
nonequilibrium stage of the collisions. The accuracy requirements result 
in a high computation memory consumption that rapidly increases with the collision
energy, approximately as $\propto s_{NN}$, and a long computation time, $\propto (s_{NN})^{3/2}$. 
The reason of this rapid rise is the Lorentz-contraction of incident
nuclei, as it is described in Ref. \cite{IRT06} in detail. 
On the one hand, 
the grid in the beam, Lorentz-contracted direction 
%($z$) 
should be fine enough
for a reasonable description of the longitudinal gradients of the matter.
From the practical point of
view, it is desirable to have  60 cells on the
Lorentz-contracted nuclear diameter%
\footnote{Though, only 40 cells per the
Lorentz-contracted nuclear diameter were possible to implement at 
$\sqrt{s_{NN}}>$ 10 GeV in order to confine the required memory and thus 
to complete the computation in a reasonable time.}.
On the other hand, 
to minimize a numerical diffusion in the computational scheme, 
an equal-step grid in all directions ($\Delta x : \Delta y : \Delta z =
1 : 1 : 1$) should be taken, in spite of Lorentz-contraction of incident
nuclei, which is quite strong at high energies. 
This choice makes the scheme isotropic with respect to the numerical
diffusion. However, 
it makes the grid too fine in the transverse directions
and thus results in high memory consumption. 
The need
of the equal-step grid in all directions for relativistic
hydrodynamic computations within conventional one-fluid
model was pointed out in Ref. \cite{Waldhauser:1992xf}. As it was
demonstrated there, the matter transport becomes even
acausal if this condition is strongly violated.

\begin{figure}[thb]
\includegraphics[width=0.48\textwidth]{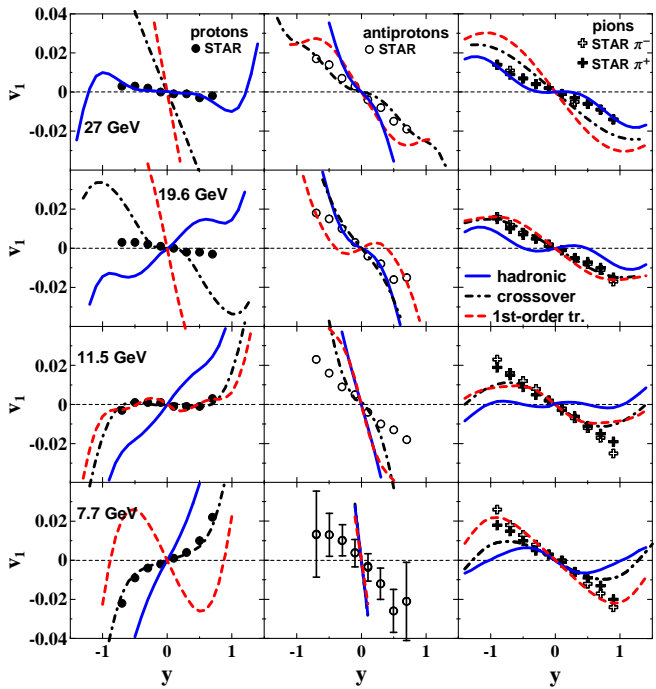}
\caption{(Color online) 
The directed flow $v_1(y)$ for protons,
  antiprotons and pions from mid-central ($b=$ 6 fm) Au+Au collisions at
  various collision energies from $\sqrt{s_{NN}}=$ 7.7 to 27 GeV
  calculated with different EoS's. Experimental data
  are from the STAR collaboration~\cite{STAR-14}.}
\label{fig:RHIC}
\end{figure}

In the simulations the acceptance
$p_T <$ 2 GeV/c for transverse momentum ($p_T$) of the produced particles 
is applied to all hadrons at all considered incident energies. 
In the 3FD model, particles are not isotopically
distinguished; i.e., the model deals with nucleons, pions,
etc. rather than with protons, neutrons, $\pi^+$, $\pi^-$ and $\pi^0$. 
Therefore, the $v_1$ values of protons, antiprotons and pions presented below, 
in fact, are $v_1$ of nucleons, antinucleons and all 
(i.e. $\pi^+$, $\pi^-$ and $\pi^0$) pions. 
Simulations are performed for mid-central collisions: impact parameter 
$b=$ 6 fm for Au+Au collisions and $b=$ 6.5 fm for Pb+Pb collisions.

The directed flow $v_1(y)$ as a function of rapidity $y$ at BES-RHIC
bombarding energies is presented in Fig.~\ref{fig:RHIC} for
pions, protons and antiprotons. 
As seen, the 3FD model does not
perfectly describe the $v_1(y)$ distributions. However, we can
definitely conclude that the best overall reproduction of the STAR data is 
achieved with the crossover EoS.
The first-order-transition scenario gives results which 
strongly differ from those in the crossover scenario, especially for the proton $v_1$. 
This is in contrast to 
other bulk observable analyzed so far 
\cite{Iv13-alt1,Iv13-alt2,Iv13-alt3,Ivanov:2012bh,Ivanov:2013cba,Ivanov:2013mxa,Iv14}.

At $\sqrt{s_{NN}}\leq$  20 GeV the the crossover EoS is certainly the best in 
reproduction of the proton $v_1(y)$. 
However, surprisingly the hadronic scenario becomes preferable for the proton $v_1(y)$
at $\sqrt{s_{NN}}>$  20 GeV.  
A similar situation takes place in the PHSD/HSD transport approach.
Indeed, predictions of the HSD model   
(i.e. without a deconfinement transition) for the proton $v_1(y)$ become preferable
at $\sqrt{s_{NN}}>$  30 GeV \cite{Konchakovski:2014gda}, 
i.e. at somewhat higher energies than in the 3FD model. 
Though, the difference between the PHSD and HSD results for protons is small 
at these energies. 
Moreover, the proton $v_1$ predicted by the UrQMD model, as cited in the experimental
paper \cite{STAR-14} and in the recent theoretical work \cite{SAP14}, 
better reproduces the proton $v_1(y)$ data at high collision energies 
than the PHSD and 3FD-deconfinement models do. 
Note that the UrQMD model is based on the hadronic dynamics. 
All these observations could be considered as an evidence of a 
problem in the QGP sector of a EoS. However the pion and antiproton $v_1$ 
contradict such a conclusion. 
Indeed, the pion and antiproton directed flow 
definitely indicates a preference of the crossover scenario at the same 
collision energies within both the  PHSD/HSD and 3FD approaches.

This puzzle has a natural resolution within the 3FD model.  
The the QGP sector of the EoS's with deconfinement  \cite{KRST06} was fitted to the lattice QCD 
data at zero net-baryon density and just extrapolated to nonzero baryon densities. 
The protons mainly originate from baryon-rich fluids that are governed by the EoS at finite baryon densities.
The too strong antiflow at $\sqrt{s_{NN}}=$ 27 GeV may be a sign of too soft QGP EoS.   
Note that a weak  
flow or antiflow indicates softness of a EoS. 
Predictions of the first-order-transition EoS, 
the QGP sector of which is constructed in the same way as that of the crossover one, 
fail even at lower collision energies,
when the QGP starts to dominate in the collision dynamics, i.e. at $\sqrt{s_{NN}}\gsim$ 15 GeV.   
This fact indirectly supports the conjecture on a too soft QGP sector 
at high baryon densities in the used EoS's.
At the same time, 
the baryon-free (fireball) fluid is governed by the EoS at zero net-baryon density.  
This fluid is a main source of antiprotons 
($\sim 80\%$ near midrapidity at $\sqrt{s_{NN}}=$ 27 GeV and $b=$ 6 fm), 
the directed flow of which is in good agreement with the data at
$\sqrt{s_{NN}}=$ 27 GeV within the crossover scenario, and in a reasonable agreement even within the 
first-order-transition scenario. 
It is encouraging because at zero net-baryon density the QGP sector of the EoS's is fitted to the lattice QCD 
data. 
The pions are produced from all fluids:  near midrapidity 
$\sim 50\%$ from the baryon-rich fluids and $\sim 50\%$ from the baryon-free one at $\sqrt{s_{NN}}=$ 27 GeV. 
Hence, the disagreement of the pion $v_1$ with data is quite moderate at $\sqrt{s_{NN}}=$ 27 GeV.

As seen from Fig.~\ref{fig:RHIC}, the deconfinement scenarios are definitely 
preferable for the pion $v_1(y)$. The pion data at 7.7 GeV are slightly better 
reproduced within the first-order-transition scenario, while at 27 GeV the crossover 
scenario is preferable. The antiproton $v_1(y)$ data testify in favor of the 
crossover scenario, except for the energy of 7.7 GeV, where all scenarios equally fail. 
It should be taken into account that the 
antiproton multiplicity in the mid-central ($b=$ 6 fm) Au+Au collision
at 7.7 Gev is $~1$ within the deconfinement scenarios and $~3$ within the hadronic scenario. 
%at b=6 fm and 11.5 Gev ~3-4 within the deconfinement scenarios and ~7 within hadronic scenario
Therefore, the hydrodynamical approach based on the grand canonical ensemble is definitely 
inapplicable to the antiprotons in this case.

\begin{figure}[thb]
\includegraphics[width=0.48\textwidth]{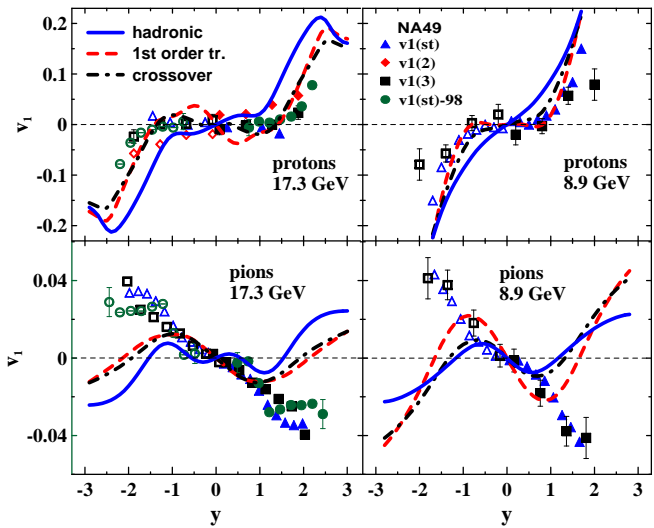}
\caption{(Color online) 
The directed flow $v_1(y)$ for protons and 
  pions from mid-central ($b=$ 6.5 fm) Pb+Pb collisions at
  collision energies $\sqrt{s_{NN}}=$ 8.9 and 17.3 GeV
  calculated with different EoS's.  
 Experimental data of the NA49 Collaboration \cite{NA49} obtained by two different methods are
displayed: the standard method [v(st)] and the method of n-particle
correlations [v(n)]. Solid symbols correspond to measured data,while
open symbols are those reflected with respect to the midrapidity. Updated
data of the NA49 Collaboration \cite{Appelshauser:1997dg} [v(st)-98] are also shown. 
}
\label{fig:SPS}
\end{figure}

Figure \ref{fig:SPS} displays a comparison of the calculated proton and pion directed 
flow from mid-central Pb+Pb collisions with the NA49 data \cite{NA49} obtained at the SPS. 
The comparison definitely  
testify in favor of the deconfinement scenarios, though 
it is difficult to choose between  
the first-order-transition and crossover scenarios. 
While the deconfinement scenarios give a reasonable agreement with the proton $v_1(y)$ in 
the whole range of rapidities, the pion $v_1(y)$ is well reproduced only in the midrapidity 
region. The calculated pion $v_1(y)$ manifests a wiggle in the midrapidity 
region while the data are monotonous functions of $y$. 
A similar situation takes place at two lower collision energies in Fig.~\ref{fig:RHIC}. 
A probable reason
for this poor reproduction of the pion $v_1$ at peripheral rapidities
is that the hydrodynamic freeze-out
disregards shadowing of a part of the frozen-out particles  
by still hydrodynamically evolving matter. 
This mechanism was discussed in Refs. \cite{RI06,Bass:1993em}. 
The shadowing means that frozen-out
particles cannot freely propagate through the region still
occupied by the hydrodynamically evolving matter but rather
become reabsorbed into the hydrodynamic phase. 
This shadowing is especially effective at the peripheral rapidities, 
where slowly-evolving near-spectator baryon-rich blobs prevent pions
with $p_x>0$ at $y>0$ and $p_x<0$ at $y<0$ from escaping 
(here $p_x$ is the transverse momentum in the reaction plane)%
\footnote{
Conventionally, it is assumed that the projectile spectator 
is situated at positive $x$ and moves with positive rapidity, 
while the target spectator, at negative $x$ and moves with negative rapidity.
}. 
This makes 
the pion  $v_1$ slope opposite in sign to the
proton $v_1$ slope at the peripheral rapidities. The hydrodynamic freeze-out does not take into account 
this shadowing, and hence, the hydrodynamic pion and proton $v_1$ slopes 
at the peripheral rapidities 
are of the same sign.

\begin{figure}[thb]
\includegraphics[width=0.48\textwidth]{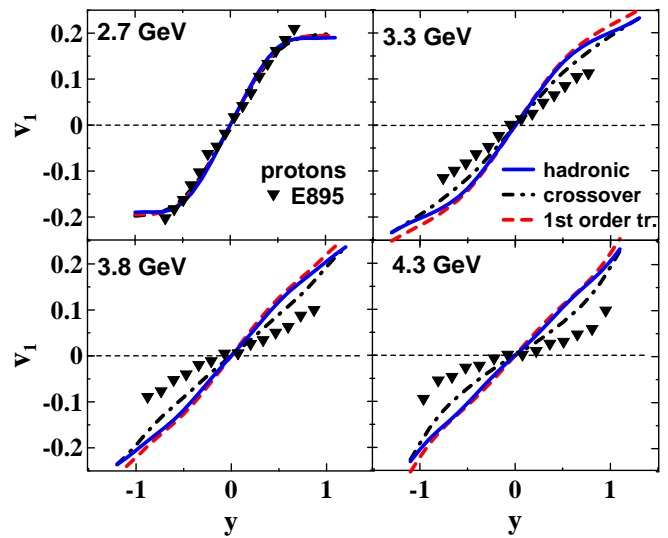}
\caption{(Color online) 
The directed flow $v_1(y)$ for protons  
  from mid-central ($b=$ 6 fm) Au+Au collisions at
  various collision energies from $\sqrt{s_{NN}}=$ 2.7 to 4.3 GeV
  calculated with different EoS's. Experimental data
  are from the E895 Collaboration~\cite{E895}.}
\label{fig:AGS}
\end{figure}
\begin{figure}[thb]
\includegraphics[width=0.48\textwidth]{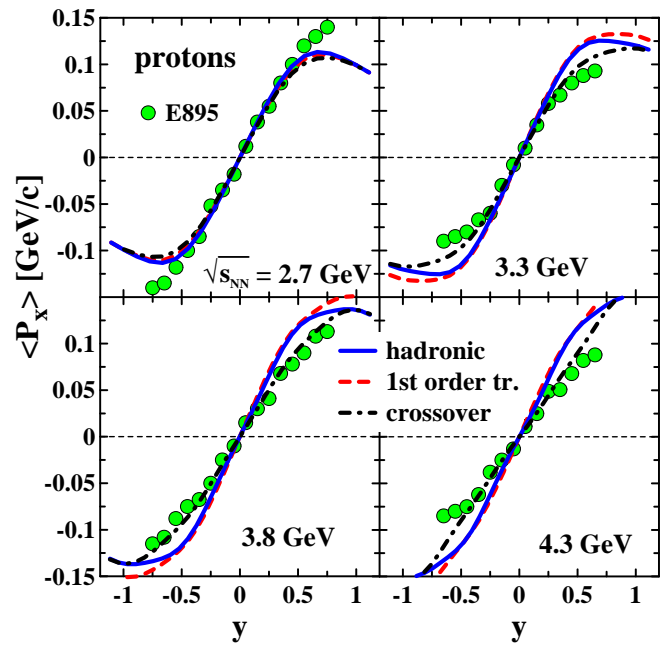}
\caption{(Color online) 
The same as in Fig. \ref{fig:AGS} but in terms of the
transverse flow $\langle P_x\rangle (y)$.}
\label{fig:AGS-Px}
\end{figure}

The directed flow $v_1(y)$ for protons  
  from mid-central Au+Au collisions at
  various collision energies from $\sqrt{s_{NN}}=$ 2.7 to 4.3 GeV
  ($E_{\rm{lab}}=$ 2, 4, 6 and 8 $A\cdot$GeV) 
  calculated with different EoS's  and its comparison with experimental data
  from the E895 collaboration~\cite{E895} 
are presented in Fig. \ref{fig:AGS}. 
As seen, at $\sqrt{s_{NN}}=$ 2.7 GeV predictions of all EoS's are identical 
(as it is expected) and are in good agreement with the data. 
With the collision energy rise the predictions of different EoS's
start to differ. First, the crossover $v_1(y)$ decouples from two others 
because the QGP fraction starts from very low densities in the crossover EoS
while the 1st-order phase transition in the corresponding EoS is not reached yet. 
At the same time, the agreement with the data worsens 
with the collision energy rise. 
The QGP fraction moves the crossover $v_1(y)$ closer to the data, though  
insufficiently close.

\begin{figure}[thb]
\includegraphics[width=7.9cm]{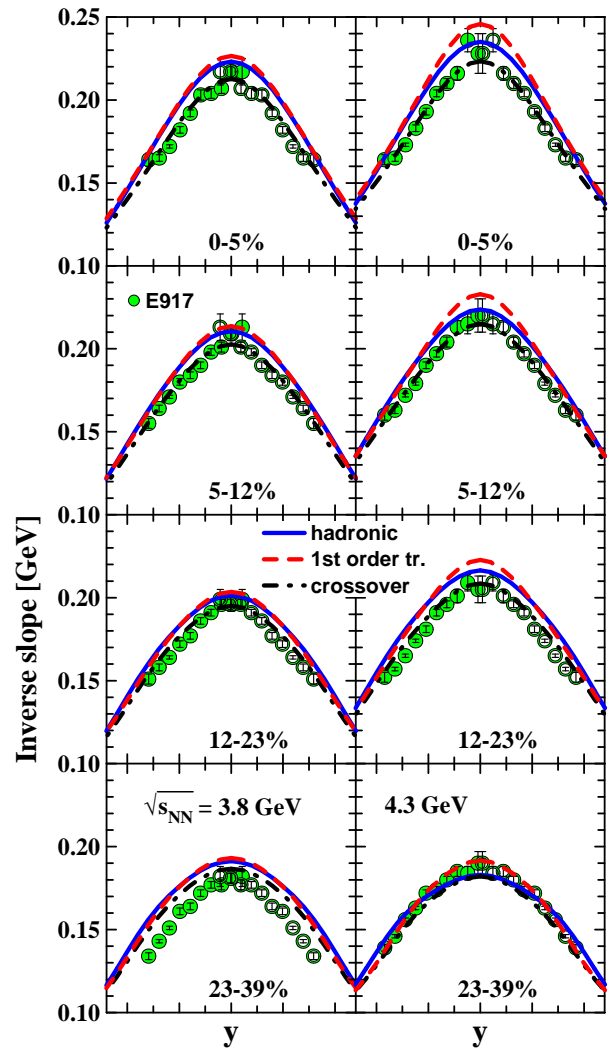}
\caption{(Color online)
Inverse-slope parameter of transverse-mass spectra of
protons, cf. $T$ in Eq. (\ref{Ttr}),  produced in Au+Au  collisions at incident  
energies  $\sqrt{s_{NN}}=$ 3.8 GeV (left column) and 4.3 GeV (right column)
[$E_{\scr{lab}} =$  6$A$ and  8$A$ GeV]  at various
centralities as a function of rapidity.
The percentage indicates the centrality, i.e. the fraction of the
total reaction cross section,  
corresponding to experimental selection of events.
The 3FD results are presented for 
impact parameters $b=$ 2, 4, 6, and 8 fm 
(from top row of panels to bottom one).  
Experimental data are from E917 Collaboration \cite{Back:2002ic}. 
Solid symbols correspond to measured data, while
open symbols are those reflected with respect to the midrapidity.
}
\label{fig:pt-slope}
\end{figure}

The E895 collaboration also presented the data \cite{E895} in terms of 
the conventional transverse-momentum flow defined 
as \cite{Danielewicz:1985hn}
\begin{eqnarray}
\langle P_x\rangle (y)= 
\frac{\displaystyle \int d^2 p_T \ p_x \ E \ dN/d^3p }%
{\displaystyle \int d^2 p_T\ E \ dN/d^3p}, 
\label{eq-px}
\end{eqnarray}
where $p_x$ is the transverse momentum of  in the reaction
plane, $E \ dN/d^3p$ is the invariant momentum distribution of a particle
with $E$ being the particle  energy, 
and integration runs over the transverse momentum $p_T$. 
These data together with results of the 3FD simulations are presented in 
Fig. \ref{fig:AGS-Px}. It is surprising that agreement of the $\langle P_x\rangle (y)$
calculated within all scenarios is much better than that in terms of $v_1(y)$. 
The crossover $\langle P_x\rangle (y)$ almost perfectly reproduces the data at all 
AGS energies. Transverse-momentum spectra of protons, which are required for recalculations 
of $\langle P_x\rangle (y)$ into $v_1(y)$, are reasonably well reproduced within all scenarios 
in the considered energy range \cite{Iv13-alt3,Ivanov:2008xx}. 
Figure \ref{fig:pt-slope} illustrates the reproduction of 
inverse-slope parameters of transverse-mass spectra of
protons, which are the only quantities that are relevant to calculations 
of $v_1(y)$ and $\langle P_x\rangle (y)$. 
These inverse slopes $T$
were deduced from fitting both the experimental \cite{Back:2002ic} and calculated proton spectra by the formula
\begin{eqnarray}
\label{Ttr}
\frac{d^2 N}{m_T \; d m_T \; d y} \propto
m_T
\exp \left(-\frac{m_T}{T}  \right),
\end{eqnarray}
where $m_T=\sqrt{m^2+p_T^2}$ and $y$ are the transverse mass and rapidity,
respectively. 
Figure \ref{fig:pt-slope} presents the calculated inverse-slope parameters of protons
produced in Au+Au  collisions at incident  
energies  $\sqrt{s_{NN}}=$ 3.8 and 4.3 GeV 
at various centralities as a function of rapidity
and their comparison with the experimental data of the  E917 Collaboration \cite{Back:2002ic}. 
As seen, the agreement with the data is indeed good, especially for the crossover EoS. 
In view of this agreement with the data on the transverse-mass spectra, 
it is puzzling that the degrees of reproduction of the
$\langle P_x\rangle (y)$ and $v_1(y)$ data are so different.

%______________________________________________________________________
\section{Midrapidity Slope of Directed Flow}
\label{Slope}

\begin{figure}[thb]
\includegraphics[width=0.45\textwidth]{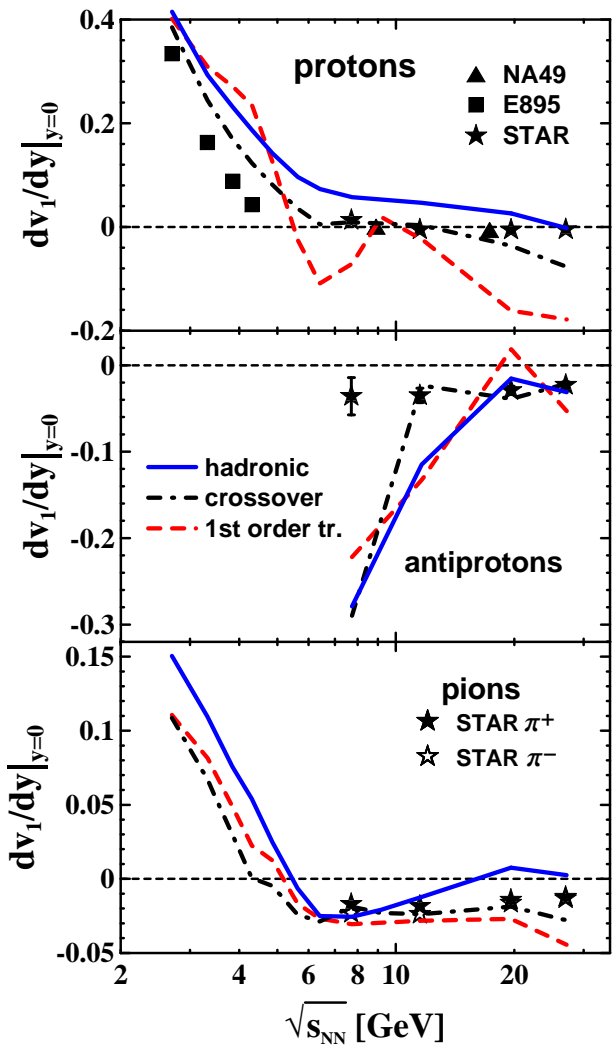}
\caption{(Color online) The beam energy dependence of the directed
  flow slope at midrapidity for protons, antiprotons and pions 
  from mid-central ($b=$ 6 fm) Au+Au collisions calculated
  with different EoS's.
   The experimental data are from the STAR measurements~\cite{STAR-14} and
  prior experiments with comparable acceptance
  cuts~\cite{NA49,E895}.}
\label{fig:slope}
\end{figure}

The slope of the directed flow at the midrapidity is often used to quantify variation 
of the directed flow with collision energy. 
The excitation functions  for the slopes of the $v_1$ distributions
at  midrapidity are presented in Fig.~\ref{fig:slope}. As noted
above, the best reproduction of the data is achieved with the 
crossover EoS. The proton $d v_1/dy$ within the first-order-transition scenario 
exhibits a wiggle earlier predicted in Refs. \cite{Ri95,Ri96,Br00,St05}.
In the present case the wiggle is mostly located in the negative range 
of slopes. The first-order-transition results demonstrate the worst 
agreement with the proton and antiproton data on $d v_1/dy$. 
The discrepancies between experiment and the 3FD 
predictions are smaller for the purely hadronic EoS, however, 
the agreement with the 3FD model for the crossover EoS is definitely better
though it is far from being perfect. All the above discussed problems 
of the crossover scenario 
at low and high collision energies reveal themselves in the $d v_1/dy$ plot.

The crossover scenario well reproduces the pion slopes at all measured energies and the 
antiproton slopes at $\sqrt{s_{NN}}>$ 10 GeV. Note that the 3FD model is poorly applicable 
to description of antiprotons  at $\sqrt{s_{NN}}<$ 10 GeV in view of a low antiproton multiplicity. 
In the case of proton slopes the crossover scenario fails both at low ($\sqrt{s_{NN}}<$ 5 GeV)
and high ($\sqrt{s_{NN}}>$ 20 GeV) collision energies. The failure at low energies is still 
questionable because the same data but in terms of $\langle P_x\rangle$ are almost perfectly 
reproduced by the crossover scenario, see Fig. \ref{fig:AGS-Px}. As for the highest 
computed energy of $\sqrt{s_{NN}}=$ 27 GeV, this simulation has been performed at the
edge of computational abilities of the code. 
%, the computation for a single impact parameter 
%takes approximately three weeks. 
Therefore, the disagreement with the data could, at least partially, a consequence of that
the 27-GeV results still suffer from insufficient accuracy. 
In the present paper the accuracy of the 27-GeV computation is slightly higher 
that in Ref. \cite{Konchakovski:2014gda}: 40 cells per the Lorentz-contracted nuclear diameter 
instead of 35 in Ref. \cite{Konchakovski:2014gda}. This made the calculated $d v_1/dy$ 
slightly closer to the experimental value: -0.077 instead of -0.086  in Ref. 
\cite{Konchakovski:2014gda}, which however is still far from the experimental value (-0.0048).

%______________________________________________________________________
\section{Conclusions}
\label{sec:conclusions}

In this study the 3FD approach has
been applied for the analysis of the recent STAR data on the
directed flow of identified hadrons~\cite{STAR-14} 
together with earlier  experimental data obtained at the SPS \cite{NA49} 
and  AGS \cite{E895}.
Simulations have 
been done with a purely hadronic EoS \cite{GM79} and two versions of the EoS
involving deconfinement transitions \cite{KRST06}---% 
a first-order phase transition and a smooth crossover transition---%
in the range of incident energies  $\sqrt{s_{NN}}$ = 2.7--27 GeV. 
Because of stringent requirements on the accuracy of the calculations we failed 
to perform calculations for energies above $\sqrt{s_{NN}}=$ 30 GeV.
The physical input
of the present 3FD calculations is described in detail in
Ref.~\cite{Iv13-alt1}. No tuning (or change) of physical 3FD-model parameters
and used EoS's
has been done in the present study as compared to that stated in
Ref.~\cite{Iv13-alt1}.

It was found that the proton directed flow within the deconfinement scenarios 
indeed manifests an antiflow (i.e. a negative slope of the $v_1$ distribution 
at the midrapidity), as it was predicted in Refs. \cite{CR99,Br00}.  
This antiflow is tiny for the crossover EoS,
which is in agreement with the data, and quite 
substantial for the EoS with the first-order phase transition. 
In the hadronic scenario, the midrapidity slope is always positive, 
except for the highest considered energy of $\sqrt{s_{NN}}$ = 27 GeV
at which a tiny antiflow is observed.
Note that the negative slope at midrapidity does
not necessarily assume a QGP EoS.  A combination
of space-momentum correlations may result in a
negative midrapidity slope of the directed flow
in high-energy nucleus-nucleus collisions  \cite{SSV00}.

The excitation function of the slope of the $v_1$ distribution 
at the midrapidity for protons turns out to
be a smooth function of the bombarding energy  without "wiggle-like" peculiarities
within the hadronic and crossover scenarios.  
At the same time, within the first-order-transition scenario this excitation function 
exhibits a wiggle earlier predicted in Ref. \cite{Ri95,Ri96,Br00,St05}.
In the present case the wiggle is mostly located in the negative range 
of slopes. The first-order-transition results demonstrate the worst 
agreement with the proton and antiproton data on the directed flow.

A high sensitivity of the directed flow, especially the proton one, to the nuclear EoS is found. 
Comparison of  other bulk observables, analyzed so far 
 \cite{Iv13-alt1,Iv13-alt2,Iv13-alt3,Ivanov:2012bh,Ivanov:2013cba,Ivanov:2013mxa,Iv14},
 with available data indicated a definite advantage of the deconfinement
(crossover and first-order) scenarios over the
purely hadronic one especially at high (RHIC) collision energies.
However, predictions of the crossover and first-order-transition
scenarios looked very similar so far. Only
a slight preference could be given to the crossover EoS.
In the case of the directed flow 
we can definitely conclude that the best overall
reproduction of the STAR data is achieved with the
crossover EoS.
%, though this reproduction is not perfect. 
The first-order-transition scenario gives
results which strongly differ from those in the crossover
scenario, especially for the proton $v_1$.

The crossover scenario well reproduces the pion $v_1$ at all measured energies and the 
antiproton flow at $\sqrt{s_{NN}}>$ 10 GeV. Note that the 3FD model is poorly applicable 
to description of antiprotons  at $\sqrt{s_{NN}}<$ 10 GeV in view of a low antiproton multiplicity. 
In the case of proton slopes the crossover scenario fails both at low ($\sqrt{s_{NN}}<$ 5 GeV)
and high ($\sqrt{s_{NN}}>$ 20 GeV) collision energies. The failure at low energies is still 
questionable because the same data but in terms of $\langle P_x\rangle$ are almost perfectly 
reproduced by the crossover scenario. As for the highest 
computed energy of $\sqrt{s_{NN}}=$ 27 GeV, this simulation has been performed at the
edge of computational abilities of the code. 
Therefore, the disagreement with the data could be, at least partially, caused by that 
the 27-GeV results still suffer from insufficient accuracy. 
On the other hand, this disagreement may indicate a problem in the QGP sector of the used crossover EoS.

The the QGP sector of the EoS's with deconfinement  \cite{KRST06} was fitted to the lattice QCD 
data at zero net-baryon density and just extrapolated to nonzero baryon densities. 
The comparison with $v_1$ data indicates that this QGP EoS at finite baryon densities 
is too soft, while the same EoS at zero net-baryon density, fitted to the lattice QCD 
data, is quite appropriate. 
Indeed, within the 3FD model 
the baryon-free (fireball) fluid is governed by the EoS at zero net-baryon density. 
This fluid is a main source of antiprotons, the directed flow of which is in good agreement with the data at
$\sqrt{s_{NN}}=$ 27 GeV within the crossover scenario, and in a reasonable agreement even within the 
first-order-transition scenario. 
The protons mainly originate from baryon-rich fluids which are governed by the EoS at finite baryon densities.
The too strong antiflow at $\sqrt{s_{NN}}=$ 27 GeV within the crossover scenario
is a sign of too soft QGP EoS.   
Predictions of the first-order-transition EoS 
%the QGP sector of which is constructed in the same way as that of the crossover one, 
fail even at lower collision energies, i.e. right above the wiggle in the excitation function 
of the proton $v_1$ slope, 
when the QGP starts to dominate in the collision dynamics.   
This fact indirectly supports the conjecture on a too soft QGP sector 
at high baryon densities in the used EoS's.
The pions are produced from all (baryon-rich and baryon-free) fluids. 
Hence, the disagreement of the pion $v_1$ with data is quite moderate at $\sqrt{s_{NN}}=$ 27 GeV.

Here it is appropriate to mention a discussion on the QGP EoS in astrophysics. 
In Ref. \cite{Alford:2004pf} it was demonstrated that the QGP EoS can be 
almost indistinguishable from the hadronic EoS at high baryon densities relevant to neutron stars. 
In particular, this gives a possibility to explain hybrid stars with masses up to about 2 solar masses ($M_\odot$),
%enables us to raise the 
%maximum hybrid star mass from about 1.6 to about 2 solar masses ($M_\odot$)
in such a way that ``hybrid stars masquerade as neutron stars'' \cite{Alford:2004pf}. 
The discussion of such a possibility has been revived after 
%the Shapiro delay 
measurements on two binary pulsars PSR J1614-2230 \cite{Demorest:2010bx} 
and PSR J0348+0432 \cite{Antoniadis:2013pzd} 
resulted in the pulsar masses of (1.97$\pm$0.04)$M_\odot$ and (2.01$\pm$0.04)$M_\odot$, respectively. 
QCD motivated models show that such a ``masquerade'' is possible 
\cite{Fukushima:2013rx}: if the
repulsive vector interaction is strong enough, it easily makes the QGP EoS sufficiently
hard. In other words, to explain the existence of the neutron star with $\sim 2M_\odot$, 
a substantially large vector interaction should be expected. Probably, the results on 
the proton directed flow give us another indication of a required hardening of the QGP EoS 
at high baryon densities.

% ____________________________________________________________________
\begin{acknowledgments}
Fruitful discussions with 
W. Cassing, V. P. Konchakovski,  V. D. Toneev, and D.N. Voskresensky 
are gratefully acknowledged. 
We are grateful to A.S. Khvorostukhin, V.V. Skokov, and  V.D. Toneev  for providing 
us with the tabulated first-order-phase-transition and crossover EoS's. 
The calculations were performed at the computer cluster of GSI (Darmstadt). 
This work was partially supported by grant NS-932.2014.2.
\end{acknowledgments}

%______________________________________________________________________

\end{document}